\documentclass[twocolumn,showpacs,preprintnumbers,amsmath,amssymb]{revtex4}

\usepackage{graphicx}
\usepackage{dcolumn}
\usepackage{bm}
\usepackage{latexsym}

\begin{document}

\preprint{APS/paper}

\title{Polarization of thin films of barium-strontium titanate under external electric field}

\author{ V. B. Shirokov }
\author{ S. V. Biryukov }
\author{ V. M. Mukhortov }
\affiliation{Southern Scientific Centre, Russian Academy of Sciences, Chehova 41, Rostov-on-Don, 344006, Russia }
\author{ Yu. I. Yuzyuk }
\affiliation{Department of Physics, Southern Federal University, Zorge 5, Rostov-on-Don, 344090, Russia}

\date{\today}

\begin{abstract}
The Landau theory of phase transitions of Ba$_{0.8}$Sr$_{0.2}$TiO$_3$ thin film under external electric field applied in the planar geometry is developed. The interfacial van-der-Waals field  $E_z=1.1 \times 10^8$ V/m oriented normal to the film-substrate interface was introduced in to the model calculation to explain experimentally observed behavior of the polarization as a function of planar electric field. The $E_z$ – misfit strain phase diagram of the film is constructed and discussed.
\end{abstract}


\maketitle

\section{\label{sec:level1}INTRODUCTION}

Thin films of (Ba$_{1-x}$Sr$_x$)TiO$_3$ (BST) are very attractive for practical applications in various fields of microelectronics and optoelectronics. Thin films of BST possess stable behavior in a wide temperatures range, a high dielectric permittivity and small losses that is very important for tunable microwave applications \cite{c1,c2,c3}. The structure and properties of thin films are well known to exhibit a number of deviations from those of bulk ceramics and single crystals. The change in the ferroelectric state is associated with high internal stresses present in thin films. They are due to a substrate-film lattice mismatch, considerable difference between the linear thermal expansion coefficients of the film and substrate, and spontaneous strain which appear at the paraelectric-ferroelectric transition when the heterostructure is cooled from the deposition temperature. A better understanding of correlations between various strain factors, which determine properties of thin films, is obviously required to achieve reliability of ferroelectric properties desirable for practical applications.

Interdigital electrodes and parallel-plate electrodes are two different types of electrodes used to measure dielectric properties at different frequencies. At present, many studies are focused on the influence of the film/electrode interface on the dielectric properties of thin films. In recent years, great attention was paid to interfacial "dead layer" with low permittivity between the film and electrode \cite{c4,c5,c6,c7}. These interfacial nano-layers are negligible in the case of bulk samples but may be comparable with the film thickness. Therefore, interfacial "dead layers" may seriously influence the effective dielectric response perpendicular to the thin film plane when usual plane capacitor geometry metal-dielectric-metal (MDM) is used for dielectric measurements with a small distance between top and bottom electrodes.

It is worth mention that interfacial "dead layers" are of minor influence for the in-plane direction of the external field created by planar electrodes deposited onto film surface where the gap between the electrodes is not connected with the film thickness. The influence of the "dead layer" is maximized in the MDM case and negligible when planar electrodes are used. However in the latter case one-dimensional phenomenological models may yield wrong results. Models describing a change of the polarization direction under external field similar to that developed for the spin-flop transitions in ferromagnetic materials \cite{c8} are obviously required.

In this paper we developed the phenomenological model for epitaxial Ba$_{0.8}$Sr$_{0.2}$TiO$_3$ (BST08) thin film on cubic substrate with the aim to describe polarization behavior under external electric field applied along the film plane. Utilizing the results of previous theoretical work \cite{c9}, we construct a misfit strain ( $u_m$ )- electric field ( $\bm{E}$ ) phase diagram BST08 assuming presence of two factors: misfit strain imposed by the substrate and electric field induced by the film-substrate interface. Existence of this interfacial field follows form the absence of the symmetry plane at the film-substrate interface. The physical origin of interfacial field is discussed in the last section of the paper. Diffuse character of phase transitions in ferroelectric thin films (dielectric maxima is always broad) is an indirect evidence of the presence of such field. It is worth note that phase transitions in ferroelectric materials under external filed conjugated with the order parameter exhibit quite similar diffused character.

In the frame of the proposed model, we calculated the dependence of each component of the polarization vector as function of planar field applied parallel to the film surface. Experimental measurements of polarization hysteresis loops were performed for the 40-nm-thick BST08 film deposited onto (001) MgO substrate. Using the theoretical model and the lattice parameters of this film we calculated the polarization as a function of planar field and estimated the value of the interfacial field.

\section{Phenomenological theory of polar states in ferroelectric thin film under external electric field}

Following Pertsev et al \cite{c10} the thermodynamic phenomenological potential of a thin film is constructed using the thermodynamic potential for the bulk sample. The thermodynamic phenomenological potential BST solid solutions and phase diagrams for BST films epitaxially grown onto (001) surface of a cubic substrate were recently reported \cite{c9}. According to this work, only one ferroelectric order parameter -polarization can be considered for BST08 thin film. The relevant expression for the cubic perovskite film deposited onto (001) surface of the cubic substrate is given as

\begin{equation}
\begin{gathered}
  G = a_1 (p_x^2  + p_y^2 ) + a_3 p_z^2 + a_{11} (p_x^4  + p_y^4 ) + a_{33} p_z^4 \hfill \\
  + a_{12} p_x^2 p_y^2  + a_{13} (p_x^2  + p_y^2 )p_z^2 + a_{111} (p_x^6  + p_y^6  + p_z^6 )  \hfill \\
  + a_{112} [p_x^4 (p_y^2  + p_z^2 ) + p_y^4 (p_x^2  + p_z^2 ) + p_z^4 (p_x^2  + p_y^2 )] \hfill \\
  + a_{123} p_x^2 p_y^2 p_z^2  + a_{1111} (p_x^8  + p_y^8  + p_z^8 ) \hfill \\
  + a_{1112} [p_x^6 (p_y^2  + p_z^2 ) + p_y^6 (p_x^2  + p_z^2 ) + p_z^6 (p_x^2  + p_y^2 )] \hfill \\
  + a_{1122} (p_x^4 p_y^4  + p_x^4 p_z^4  + p_y^4 p_z^4 ) \hfill \\
  + a_{1123} (p_x^4 p_y^2 p_z^2  + p_x^2 p_y^4 p_z^2  + p_x^2 p_y^2 p_z^4 ). \hfill \\
\end{gathered}
\end{equation}

All coefficients of the potential (1) were calculated for particular concentration BST08 using the coefficients of the end-members of the solid solution according to the formalism developed for BST solid solution \cite{c9}. The high-temperature asymptotic values of the coefficients in the quadratic terms of the potential for BST08 yield

\[
\begin{gathered}
a_1  = 4.791 \times 10^5 (T - 328) - 0.857 \times 10^{10} u_m \text{  }Jm /C^2,  \hfill \\
a_3  = 4.791 \times 10^5 (T - 328) + 1.014 \times 10^{10} u_m \text{  }Jm /C^2,  \hfill \\
\end{gathered}
\]

where $u_m$ is misfit strain. All calculated coefficients are listed in Table I.

\begin{table}
\caption{\label{tab1}
Coefficients of the phenomenological potential (1) for BST08 film deposited onto (001) surface of the cubic substrate}
\begin{ruledtabular}
\begin{tabular}{ccc}
Coefficient& BST08 &  Units  \\ \hline \\
$a_1$     &  $4.791(T - 328) - 0.857 \times 10^{5} u_m$   &  $ \times 10^{5} Jm /C^2$ \\
$a_3$     &  $4.791(T - 328) + 1.014 \times 10^{5} u_m$   &  \\
\\
$a_{11}$  &  1.6837   & $\times 10^{8} Jm^5/C^4$\\
$a_{12}$  &  4.1796   &   \\
$a_{13}$  &  4.4478   &   \\
$a_{33}$  &  -1.4913  &   \\
\\
$a_{111}$ &  1.0352    &  $\times 10^{9} Jm^9/C^6$\\
$a_{112}$ &  -1.5600   &  \\
$a_{123}$ &  -0.6080   &  \\
\\
$a_{1111}$ &  3.0904  &  $\times 10^{10} Jm^{13}/C^8$\\
$a_{1112}$ &  2.0232  &  \\
$a_{1122}$ &  1.3096  &  \\
$a_{1123}$ &  1.0936  &  \\

\end{tabular}
\end{ruledtabular}
\end{table}

Strain of a film with respect to the bulk BST08 sample are

\begin{equation}
\begin{gathered}
  e_1  = u_m , \hfill \\
  e_2  = u_m , \hfill \\
  e_3  =  - 0.00659p_x^2  - 0.00659p_y^2  + 0.04276p_z^2   \hfill \\
   \qquad \text{ } - 0.84668u_m , \hfill \\
  e_4  = 0.02673p_y p_z , \hfill \\
  e_5  = 0.02673p_x p_z , \hfill \\
  e_6  = 0. \hfill \\
\end{gathered}
\end{equation}

\begin{figure*}
\includegraphics{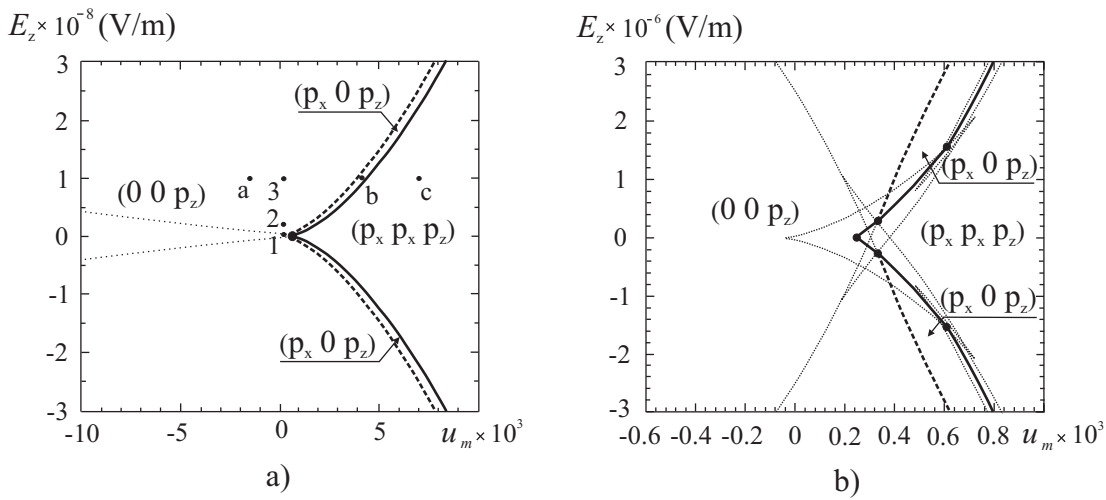}
\caption{\label{fig1}
a) Misfit strain $u_m$ - interfacial field $E_z$ phase diagram BST08 thin film. b) Enlarged sections of the diagram near $E_z=0$ and $u_m=0$. Solid and dashed lines correspond to first- and second-order phase transitions, respectively. Thin dotted lines correspond to the metastable states.}
\end{figure*}

In the following we consider the single-domain film deposited onto (001) face of the cubic substrate with $z$ axis being normal to the film surface. External electric field $E_x$ is applied using planar electrodes deposited on the film surface and aligned parallel to one of the substrate's cubic axes - $y$. The interfacial van-der-Waals field $E_z$ arises at the interface of two surfaces and depends strongly on the technological parameters of the film growth. To analyze the effect of the external electric field, additional terms  $ - E_x p_x  - E_z p_z $  should be included into potential (1). One more term responsible for the depolarizing filed should be included in the potential (1). However, in our simple model we assume that inhomogeneous polarization on the film surface equal to zero due to presence of inhomogeneous interfacial van-der-Waals field.

Fig. 1 shows the misfit strain ( $u_m$ ) - interfacial field $E_z$ phase diagram under zero external planar field ( $E_x =0$ ). A component of polarization $p_z$ normal to the film surface exists at any $u_m$. Two additional field-induced phases appear at positive $u_m$ values with additional component of the polarization along the cubic axis ($p_x$ 0 $p_z$) and along the diagonal of the film surface ($p_x$ $p_x$ $p_z$). The phase transition between ($p_x$ 0 $p_z$) and $p_x$ $p_x$ $p_z$) is of first-order type (bold line in Fig.1), while transition between (0 0 $p_z$) and ($p_x$ 0 $p_z$) is of second-order type (dashed line in Fig.1). At $E_z=0$, the phase transition between ($p_x$ $p_x$ $p_z$) and (0 0 $p_z$) (first order) occurs at $u_m=0.23 \times 10^{-3}$, however, ($p_x$ $p_x$ $p_z$) is metastable up to $u_m=-0.05 \times 10^{-3}$.

Dotted lines in Fig. 1a show a range of $180^{\circ}$ domains of the tetragonal film while above and below these lines the film is in single-domain state with alternative direction of polarization $+p_z$ and $-p_z$ , respectively. Enlarged sections of the diagram near $E_z=0$ and $u_m=0$ is shown in Fig.1b. According to this diagram, on increasing $u_m$ at $E_z=0$ the polarization changes from the direction orthogonal to the substrate (0 0 $p_z$) to the tilted one - ($p_x$ $p_x$ $p_z$). Note that polarization direction ($p_x$ 0 $p_z$) is possible at nonzero $E_z$ values.

In the following we analyze the polarization of the film as a function of external electric field $E_x$. In various parts of the phase diagram dependence of polarization is quite different especially near the phase transition lines, and very complicated around the multiphase point.

Polarization as a function of ac external electric  field   and interfacial field $E_z$ = const was calculated using dynamic equations of Landau-Khalatnikov with a uniform relaxation time for different component of the polarization.

\begin{figure}
\includegraphics{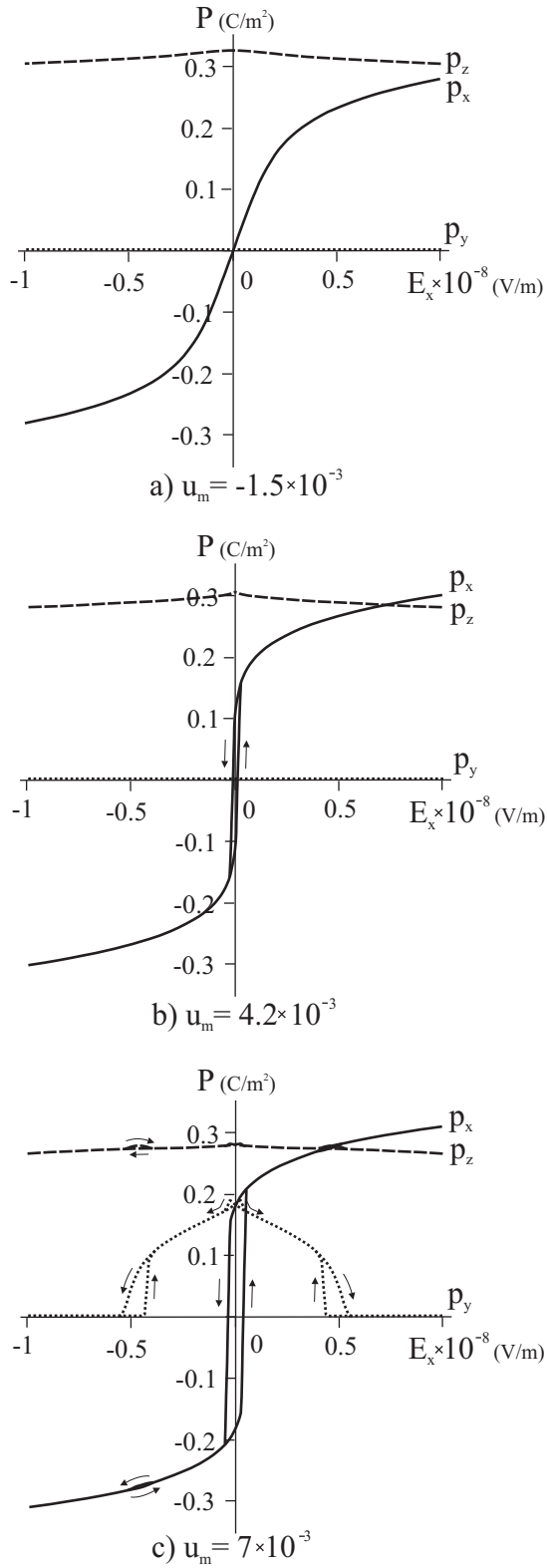}
\caption{\label{fig2}
Three components of polarization as a function of ac electric field $E_x$ under various values of misfit strain at fixed interfacial field $E_z=10^8$ V/m. A solid line - $p_x$, dotted line - $p_y$, a dashed line - $p_z$. Figures a-c corresponds to the points indicated in Fig. 1a.}
\end{figure}

\begin{equation}
\begin{gathered}
 \tau \frac{{\partial p_x }}
{{\partial t}} =  - \frac{{\partial G}}
{{\partial p_x }}, \hfill \\
  \tau \frac{{\partial p_y }}
{{\partial t}} =  - \frac{{\partial G}}
{{\partial p_y }}, \hfill \\
  \tau \frac{{\partial p_z }}
{{\partial t}} =  - \frac{{\partial G}}
{{\partial p_z }}. \hfill \\
\end{gathered}
\end{equation}

A Runge-Kutte method was employed for numerical calculations. Fig. 2 shows typical loops calculated for all components of polarization at $E_z=10^8$ V/m in several point of the phase diagram corresponding to the different values of misfit strain. The chosen value of $E_z$ used in our calculations will be justified below. Dependences of all components of polarization as a function of external electric field $E_x$ become complicated with increasing $u_m$. In the tetragonal phase (0 0 $p_z$) the $p_x$ component (solid line in Fig. 2a) exhibits nonlinear behavior. On approaching the phase transitions lines the slop of the $p_x$ component increases and hysteresis starts to appear (Fig. 2b). Finally, in the phase ($p_x$, $p_x$, $p_z$) the $p_y$ component exhibits dome-like feature around zero $E_x$ values (Fig. 2c) and hysteresis behavior appears at the edges of the dome and near zero $E_x$ values as well. It worth noting that all three components of the polarization exhibits hysteresis inside the ($p_x$, $p_x$, $p_z$) phase.

\section{Experimental results}

The transparent and mirror-smooth heteroepitaxial BST08 thin film (thickness 40 nm) was deposited on (001)MgO single crystalline 0.5 mm-thick substrate by rf sputtering of polycrystalline target of the corresponding stoichiometric composition. The temperature of the substrate was kept at $650-750^\circ$C. Details of the growth conditions have been previously reported \cite{c11}. The epitaxial relationships between the film and the MgO substrate have been confirmed by X-ray diffraction (XRD): $(001)_{\text{BST}} \parallel (001)_{\text{MgO}}$ and $[100]_{\text{BST}} \parallel  [100]_{\text{MgO}}$. XRD study revealed that the room-temperature unit cell parameter of the film normal to the substrate ($c = 0.39936$ nm) is smaller than that in plane of the film ($a = 0.40032$ nm).  At $600^\circ$Á the lattice parameters of this film were determined to be $c=0.40118$ nm and $a=0.40287$ nm.

For application of external electric field, a periodic array of 440 interdigital Al electrodes were deposited on the top of the film using standard optical photolithography. The electrodes thickness, length and width were 300, 130 and 2 $\mu$m, respectively. The spacing between the electrodes was 1 $\mu$m. The electrodes were aligned parallel to one of the substrate's cubic axes. An optical microscope "Leitz Latimet" was used to verify topology and quality of the electrodes. Capacity, conductivity and leakage current were measured at 1 kHz using Keihtley 4200SCS analyzer and PM-5 MicroTec probe station.

Room-temperature polarization-electric-field hysteresis measurements were performed in a frequency range $10^{-2}$ - $10^3$ Hz using a modified Sawyer-Tower circuit \cite{c12}.

To estimate the film strain we used the equation \cite{c13}:

\begin{equation}
u_m  = u_0  - (\alpha _f  - \frac{{a_s }}
{{a_f }}\alpha _S )(T - T_0 )
\end{equation}

\begin{figure}
\includegraphics{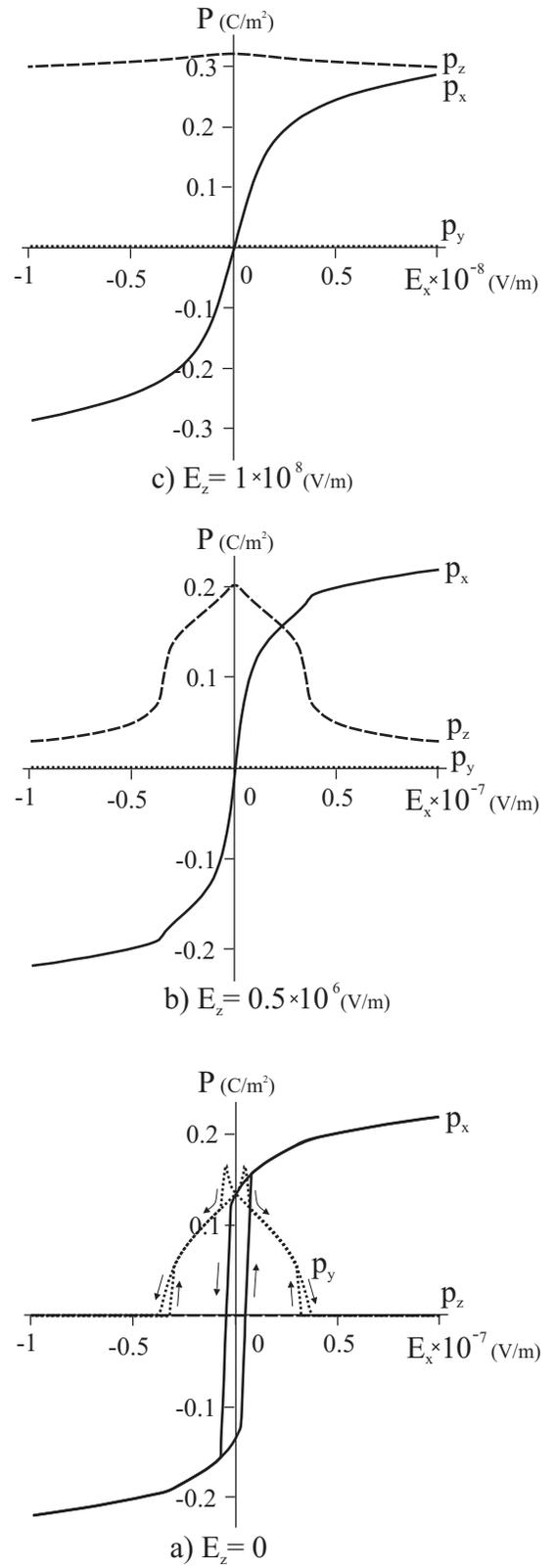}
\caption{\label{fig3}
Three components of polarization as a function of ac electric field $E_x$ under various values of a interfacial field at fixed misfit strain $u_m=0.16 \times 10^{-3}$. A solid line - $p_x$, dotted line - $p_y$, a dashed line - $p_z$. Figures a-c corresponds to the points 1-3 indicated in Fig. 1a.}
\end{figure}

\begin{figure}
\includegraphics{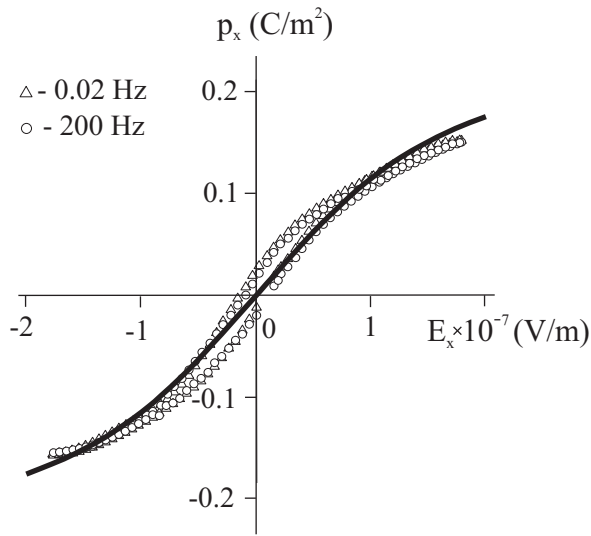}
\caption{\label{fig4}
Experimental dependence (open symbols) of $p_x$ component of the polarization as a function of ac electric field $E_x$ and theoretical dependence (bold line) of this component at $E_z=1.1 \times 10^8$ V/m and $u_m=0.16 \times 10^{-3}$.}
\end{figure}

where  $\alpha_f$ ,  $\alpha_s$  -are the thermal expansion coefficients of the film and substrate, respectively; $a_f$ and $a_s$ - are the lattice parameters of the film and substrate at $T_0$ ; $u_0$ - is initial strain at at $T_0$. At room temperature $a_s$=0.4212 and thermal expansion coefficient  $\alpha_s=1.2 \times 10^{-5}$ 1/K above room temperature up to $600^\circ$C \cite{c14}, therefore $a_s$=0.4247nm at $600^\circ$C. The lattice parameters of bulk BST08 solid solution were calculated according Eqs. (7) and (8) from Ref. \cite{c9}. Using the structural information about end-members of the solid solution \cite{c15,c16}, we found  $\alpha_f=0.922 \times 10^{-5}$ 1/K and $a_f$=0.3998 nm. Assuming that spontaneous polarization is zero at $600^\circ$C and using the relation $c/a=(1+e_3)/(1+e_1)$, Eqs. (2) and (4) at $T=T_0$ we found $u_0=2.28 \times 10^{-3}$. Therefore, the room-temperature misfit strain calculated using Eq. (4) is $u_m=0.16 \times 10^{-3}$.

Fig. 3 shows typical loops calculated for all components of polarization at fixed misfit strain $u_m=0.16$ in several points of the phase diagram corresponding to the different values of the interfacial field $E_z$. These three points are indicated by numbers 1-3 in Fig. 1a.

The polarization as a function of ac electric field $E_x$ calculated at zero interfacial field (Fig. 3a ) does not agree with the experimental loop presented in Fig. 4. Reasonable agreement can be achieved at $E_z=1.1 \times 10^8$ V/m. The corresponding theoretical dependence of $p_x$ component as a function of ac electric field $E_x$ is shown in Fig.4 (bold line).

\section{Discussion and conclusions}

A phenomenological model including three components of the polarization was developed to explain the dependence of the in-plane component of the polarization $p_x$ as a function of ac electric field $E_x$ applied parallel to the film plane. The misfit strain $u_m$ of the film was found to be equal to $0.16 \times 10^{-3}$. The interfacial van-der-Waals field $E_z=1.1 \times 10^8$ V/m oriented normal to the film-substrate interface was introduced in the model calculation to explain experimentally observed behavior of the polarization as a function of planar electric field. In the following we justify the existence of the interfacial van-der-Waals field.

A short-range van-der-Waals interaction always exits between two atoms due to mutual polarization of electronic shells. In some cases, the interaction energy of two planar surfaces with the distance $r$ is  $\sim r^{ - 2}$ for unit area of one surface interacting with an infinite area of another surface \cite{c17}. As known from the tip-surface curves in atomic-force microscopy \cite{c18} the radius of the interaction varies from several angstroms up to several tens of nanometers. These van-der-Waals fields are only partially compensated when two planar surfaces are in contact. The resulting energy and direction of this filed at the interface of two surfaces is a subject of quantum-mechanical calculations. In fact, the interfacial van-der-Waals field is a result of the polarization induced at the substrate.

If polarization at the external film surface is not equal to zero, depolarizing field appears to reduce the spontaneous polarization \cite{c19}. The direction of the depolarizing field is opposite to the direction of the interfacial van-der-Waals field, therefore the depolarizing field reduces the effect of the interfacial one. The interfacial van-der-Waals field is a long-range inhomogeneous field \cite{c17}, therefore our homogeneous phenomenological model should be considered as approximation of the averaged inhomogeneous model in which boundary conditions on the external surface of the film allows absence of the depolarizing field.

In general case, the interfacial van-der-Waals field fluctuates and its direction is not defined. However, under external field the van-der-Waals field acquires particular direction since fluctuating dipoles become oriented under external field. Therefore, below the ferroelectric phase transition temperature, the interfacial van-der-Waals field has preferred direction normal to the substrate.

In fact, the introduction of the interfacial van-der-Waals field is equivalent to the introduction of so-called "passive layer" \cite{c20}. Very likely, that origin of the "passive layer" is the effect of the van-der-Waals fields in these (the interfacial) regions. The size of the "passive layer" depends on the physical properties of the contacting materials. Namely, the field strength and its long-range properties depend mainly on the spectral characteristics of the dielectric constants of the contacting materials \cite{c21,c22}.

The calculated $p_x(E_x)$ dependence (bold line in Fig. 4) implies that experimentally observed losses (loops in Fig. 4) appear due to relaxations in the regions close to electrodes and not due to the losses inside the material of the film studied. Therefore, in functional devices based on this film one can expect low losses in microwave frequency range. As follows from Fig. 2, by increasing the misfit strain one can increase the slope of the hysteresis loop and therefore increase the tunability. However, large positive misfit strains induce the hysteresis behavior implying large losses inside the material of the film.

\begin{acknowledgments}

This study was supported by the Russian Foundation for Basic Research under Projects Nos 09-02-00254a and 09-02-00666a.

\end{acknowledgments}


\end{document}